\documentclass[aps, pra, twocolumn, superscriptaddress, amsmath, 
tightenlines, longbibliography]{revtex4-2}

\usepackage{amssymb}
\usepackage{amsmath}
\usepackage{dcolumn}
\usepackage{graphicx}
\usepackage{mathrsfs}
\usepackage{appendix}
\usepackage{graphicx}
\usepackage{booktabs}
\usepackage{units}

\def \hide#1{}

\usepackage{url}
\usepackage[colorlinks]{hyperref}
\hypersetup{%
	plainpages=true,
	breaklinks=true,       %not default in dvips mode, so we must specify
	hypertexnames=false,  %not ideal, but needed when pagenums duplicate (`i' vs. 
	pageanchor=true,
	colorlinks=true,
	linkcolor={blue},
	citecolor={red},
	urlcolor={blue},
	anchorcolor={black}
}

\begin{document}

	\title{Selective band interaction and long-range hopping in a structured environment with giant atoms }

	\author{Ying Xia}
	\affiliation{Institute of Theoretical Physics, School of Physics, Xi'an Jiaotong University, Xi'an 710049, People’s Republic of China}
	
	\author{Jia-Qi Li}
	\affiliation{Institute of Theoretical Physics, School of Physics, Xi'an Jiaotong University, Xi'an 710049, People’s Republic of China}
	
	\author{Xin Wang}\email{wangxin.phy@xjtu.edu.cn}
	\affiliation{Institute of Theoretical Physics, School of Physics, Xi'an Jiaotong University, Xi'an 710049, People’s Republic of China}

\date{\today}

\begin{abstract}
Giant atoms, which couple to the environment at multiple discrete points, exhibit various nontrivial phenomena in quantum optics due to their nonlocal couplings. In this study, we propose a one-dimensional cross-stitch ladder lattice featuring both a dispersive band and a flat band. By modulating the relative phase between the coupling points, the giant atom selectively interacts with either band. First, we analyze the scenario where the dispersive and flat bands intersect at two points, and the atomic frequency lies within the band. Unlike the small atom, which simultaneously interacts with both bands, a single giant atom with a controllable phase interacts exclusively with the dispersive or flat band.  Second, in the bandgap regime, where two atoms interact through bound-state overlaps manifesting as dipole-dipole interactions, we demonstrate that giant atoms enable deterministic long-range hopping and energy exchange with higher fidelity compared to small atoms. These findings provide promising applications in quantum information processing, offering enhanced controllability and selectivity for quantum systems and devices.
\end{abstract}

\maketitle

\section{introduction}
A fundamental goal in quantum optics is to precisely control light–matter interactions at the single-photon level~\cite{Cohen1998}, particularly in systems where quantum emitters couple to structured environments with tailored spectral properties~\cite{Wang2022Oct,Lambropoulos2000,Gonzalez-Tudela2017,Leonforte2024}. When the atomic transition frequency matches the center of a spectral band, resonant coupling enhances spontaneous emission into guided modes~\cite{Scully1997,Bykov1975,Stewart2020,Mirhosseini2019}. This is closely related to the Purcell effect, which boosts the emission rate by increasing the coupling between the emitter and the photonic environment~\cite{Purcell1995,Rybin2016,Wang2021}. Conversely, when atomic frequencies are tuned within photonic band gaps—regions devoid of propagating modes—bound states form, effectively trapping atomic excitation and preventing decay, thus facilitating long-lived coherence~\cite{John1990,WangX2021,Liu2016,Plotnik2011,Vega2021}. Such states are essential for applications in quantum memory and delayed photon emission~\cite{John1994,Kurizki1990,RomnRoche2020}. Additionally, structured environments support phenomena like subradiance and superradiance~\cite{Cardenas-Lopez2023,Dicke1954,Gross1982,Sinha2020,Svidzinsky2008,Slepyan2013,Jenkins2017,Ke2019,Zhang2019,Bienaim2012}, where collective interference either suppresses or enhances emission rates. 

In lattice systems, specific geometric configurations can induce destructive interference~\cite{Leykam2018,Derzhko2015,BERGHOLTZ2013,Leykam2018x}, resulting in dispersionless flat bands characterized by compact localized states~\cite{Sutherland1986,Benedetto2024,Aoki1996,Johansson2015,Miyahara2005,Hyrks2013,MoralesInostroza2016}. These flat bands give rise to unique phenomena such as caging effects~\cite{Vidal1998c,Danieli2021v,Danieli2021b}, unconventional Anderson localization~\cite{Goda2006,Longhi2021,Anderson1958m}, and superconductivity~\cite{Aoki2020,Kopnin2011a,Iglovikov2014b,Cao2018s,Mondaini2018f}. When an emitter couples to a flat band, destructive interference suppresses propagation modes, trapping excitations into non-radiative dark states that prevent energy dissipation and extend coherence~\cite{Vidal1998,Martinez2023g}. These unique interactions within engineered photonic lattices enable quantum memory and robust photon-mediated entanglement, paving the way for advanced applications in integrated photonic circuits and quantum information technologies.

Giant emitters, which interact with the bath at multiple spatially separated points, breaking the dipole approximation, are different from the point-like small atom~\cite{Kannan2020,Cai2021,TerradasBrians2022,Chen2023,Zhang2022}. This nonlocal coupling leads to distinctive interference effects which alter the atomic decay dynamics and give rise to novel quantum phenomena beyond small atoms. For instance, giant atoms can exhibit both frequency-dependent decay rates~\cite{FriskKockum2014,Du2022,Du2022x} and phase-controlled interference effects~\cite{Du2022y}, enabling chiral quantum optics~\cite{Joshi2023,Wang2022x,Zhou2023,Soro2022,Wang2021y}. Moreover, in structured lattice systems with various bands, small atoms interact with all modes if they resonate at the cross-points of energy bands. However, by tuning the relative phase between different coupling points, giant atoms can selectively interact with specific modes, allowing for phenomena like non-Markovian dynamics~\cite{Roccati2024,Guo2017,Qiu2023,Yin2022,Andersson2019x,Guo2020x} and the formation of bound states in the continuum~\cite{Soro2023,Guo2020c,Xiao2022v,Zhao2020b}. 

In this work, we investigate the selective interaction properties of giant atoms in a 1D cross-stitch lattice comprising flat and dispersive bands. By tuning their relative positions, we explore the dynamical evolution of small and giant emitters, highlighting distinct interactions with localized and propagating modes. Firstly, we derive the spectra of the lattice and an effective model after transformation. Secondly, we examine the case where the dispersive and flat bands intersect, showing that while small atoms interact with both bands, giant atoms selectively couple to either band depending on the relative phase. Thirdly, when a bandgap separates the flat and dispersive bands, bound states form if the atomic frequency lies within the gap. We analyze the dipole-dipole interaction between two separated atoms, demonstrating that giant atoms achieve high-fidelity interactions due to their selective coupling. 

\section{Spectrum of 1D cross-stitch lattice}
\begin{figure*}[ht]
\centering \includegraphics[width=\textwidth]{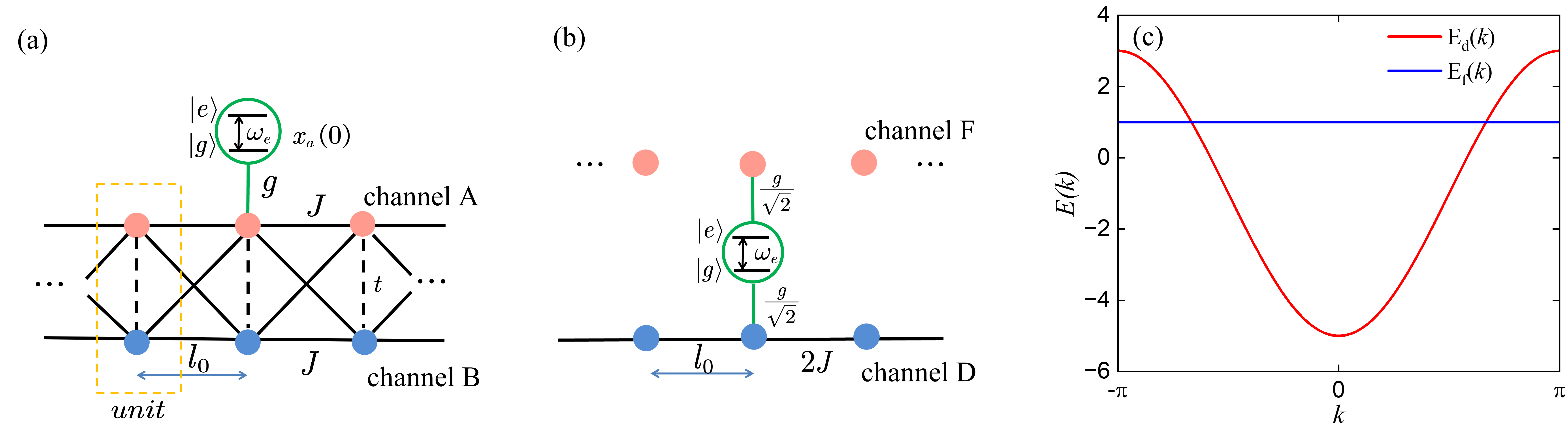}
\caption{(a) A two-level emitter interacts with the 1D cross-stitch lattice structure. Each unit cell with two sublattices ($A$ and $B$) is shown in the orange dashed box. $J$ and $t$ are the inter-cell  and intra-cell hopping amplitudes. (b) The equivalent lattice model after transformation. The hopping amplitude between the nearest-neighboring sites of the channel D is $2J$. (c) The band structure of the 1D cross-stitch model, featuring a flat band and a dispersive band. The parameters are $J=-1$, $t=1$.}
\label{fig1}
\end{figure*}

As illustrated in Fig.~\ref{fig1}(a), we consider a two-level quantum emitter with a frequency $\omega_e$ interacting with an artificial quasi-1D cross-stitch lattice model. The lattice consists of $N$ unit cells (with $2N$ lattice sites). Each unit cell, outlined by the orange dashed box, comprises two sublattices $A$ and $B$. The black dashed lines indicate intra-cell hopping $t$, while the black solid lines represent inter-cell hopping $J$. For simplicity, we set the length of a single unit cell as $l_0=1$. The tight-binding Hamiltonian is ($\hbar=1$)
\begin{align}
	H_L &= \sum_x \omega_0 \left( a_{x}^{\dagger} a_x + b_{x}^{\dagger} b_x \right) - \left[ \sum_x t a_{x}^{\dagger} b_x \right. \nonumber \\
	&\quad + \left. \sum_x J \left( a_{x}^{\dagger} + b_{x}^{\dagger} \right) \left( a_{x+1} + b_{x+1} \right) + \mathrm{H.c.} \right],
\end{align}
where  $\omega_0$ is the identical frequency of the bosonic modes and $a_x$ ($b_x$) is the photon annihilation operator of the sub-sites $a$ ($b$) at the $x$-th unit. In the following, we derive the Hamiltonian in the rotating frame of atomic frequency $\omega_0$.

Via inverse Fourier transformation, the real space operator is rewritten in momentum space
\begin{equation}
	a_x=\frac{1}{\sqrt{N}}\sum_k{e^{ikx}a_k},\quad b_x=\frac{1}{\sqrt{N}}\sum_k{e^{ikx}b_k},
\end{equation}
where $	k=2\pi n/N,\ n\in \left( -N/2,N/2 \right]$. The Hamiltonian is expressed in $k$ space
\begin{align}
H_L&=\left( \begin{matrix}
	a_{k}^{\dagger}&		b_{k}^{\dagger}\\
\end{matrix} \right) \mathcal{H} _L\left( \begin{array}{c}
	a_k\\
	b_k\\
\end{array} \right), 
\\
\mathcal{H} _L&=\left( \begin{matrix}
	-2J\cos k&		-2J\cos k-t\\
	-2J\cos k-t&		-2J\cos k\\
\end{matrix} \right) \nonumber
\\
&=\left( -2J\cos k \right) I+\left( -2J\cos k-t \right) \sigma _x.
\end{align}
Note that $\mathcal{H} _L$  is expressed in terms of Pauli operators, indicating that the $A$ and $B$ sites behave as an effective spin system. This effective spin is not independent but is coupled through the inter-cell hopping $J$ and intra-cell hopping $t$. The dispersion relations and eigenmodes can be straightforwardly obtained by diagonalizing $\mathcal{H} _L$
\begin{align}
	E_f&=t,\quad E_{kd}=-4J\cos k-t,\\
	C_{k,f}^{\dagger}&=\frac{1}{\sqrt{2}}\left( \begin{matrix}
		a_{k}^{\dagger}&		-b_{k}^{\dagger}\\
	\end{matrix} \right),\quad	C_{k,d}^{\dagger}=\frac{1}{\sqrt{2}}\left( \begin{matrix}
		a_{k}^{\dagger}&		b_{k}^{\dagger}\\
	\end{matrix} \right). 
\end{align}

\noindent Here, the subscripts $f$ and $d$ denote the flat and dispersive bands, respectively. Note that $C_{k,f}^{\dagger}$ and $C_{k,d}^{\dagger}$ are superpositions of operators $a_{k}^{\dagger}$ and $b_{k}^{\dagger}$, which are decoupled.
The the explicit unitary transformation between the models is $H_{L}=U^{-1}H_LU$
\begin{align}
\left( \begin{array}{c}
			C_{x,f}\\
			C_{x,d}\\
		\end{array} \right) =U\left( \begin{array}{c}
			a_x\\
			b_x\\
		\end{array} \right) =\frac{1}{\sqrt{2}}\left( \begin{matrix}
			1&		1\\
			-1&		1\\
		\end{matrix} \right) \left( \begin{array}{c}
			a_x\\
			b_x\\
		\end{array} \right) ,\label{transfor mattrix}
\end{align}
Consequently, we transform the lattice into an equivalent model, as shown in Fig.~\ref{fig1}(b). The Hamiltonian is
\begin{align}
	H_{L} &= t \sum_x C_{x,f}^{\dagger} C_{x,f} \nonumber\\
	&\quad - \sum_x \left( t C_{x,d}^{\dagger} C_{x,d} - 2J C_{x+1,d}^{\dagger} C_{x,d} \right) + \mathrm{H.c.}
\end{align}
The band structure of the model is depicted in Fig.~\ref{fig1}(c). Because channels of flat and dispersive bands are independent, the dispersive band is centered at $-t$, while the flat band has an energy of $t$. By tuning the parameter $t$, the energy band structure can be controlled, allowing adjustment of the relative positions of the two bands. Notably, when $|t|>|2J|$, the flat band becomes separated from the dispersive band, resulting in the emergence of a band gap.

\section{Emitter Frequency at Intersection Points of two bands}
To demonstrate the selective coupling characteristics of giant atoms, we examine two representative scenarios: the emitter's frequency lies within the energy bands, or falls within the bandgap. These cases emphasize the role of dispersive and flat bands in mediating the interaction between the atom and the lattice. We first consider $|t|<|2J|$, where the flat and dispersive bands intersect at two points. In this scenario, we analyze a single two-level emitter with frequency $\omega_e$ located within the bands, interacting with both the flat and dispersive bands.

\subsection{Non-Selective Band Interaction of a Small Atom}
Firstly, we consider that the emitter has a small atom form, coupled to the lattice at $x_a(0)$. In this situation, the system's Hamiltonian is written as
\begin{align}
	H_S&=H_0+H_{\mathrm{int}},\quad	H_0=\frac{1}{2}\omega _e\sigma _z+H_L,\label{H_s}\\
	 H_{\mathrm{int}}&=g\left( \sigma _-a_{0}^{\dagger}+\sigma _+a_0 \right),\label{eq:Hint}
\end{align}
where $\sigma _{z,\pm}$ are the Pauli operators of the emitter and $g$ is the coupling strength. Using  Eq.~(\ref{transfor mattrix}) and performing the Fourier transform on both sites, we obtain the equivalent interaction Hamiltonian
\begin{equation}
H_{\mathrm{int}}=g\sigma _- \frac{\left( C_{0,f}^{\dagger}+C_{0,d}^{\dagger} \right)}{\sqrt{2}} +\mathrm{H}.\mathrm{c}.
\end{equation}
After the transformation, the small emitter in real space becomes a giant form in the equivalent lattice space, coupling to two points. One leg couples to  $C_{0,f}$, i.e., a single-mode cavity, while the other leg couples to $C_{0,d}$, a site in a chain, as shown in Fig.~\ref{fig1}(b). A similar mapping strategy has been introduced in ~\cite{Roccati2022} to connect small and giant atom configurations. Applying the inverse Fourier transform, $H_{\mathrm{int}}$ is rewritten as
\begin{equation}	
	H_{\mathrm{int}}=\frac{g}{\sqrt{2N}}\sum_k{\sigma_{-} \left( C_{k_f}^{\dagger}+C_{k_d}^{\dagger} \right)}+\mathrm{H}.\mathrm{c}.
\end{equation}
After applying the unitary transformation $U_0\left( t \right) =\exp \left( -iH_0t \right) $, the interaction Hamiltonian is
\begin{equation}
H_{\mathrm{int}}=\frac{g}{\sqrt{2N}}\sum_k{\left( \sigma _-C_{k_f}^{\dagger}e^{i\Delta _ft}+\sigma _-C_{k_d}^{\dagger}e^{i\Delta _{kd}t} \right)}+\mathrm{H}.\mathrm{c}.,\label{eq:eff}
\end{equation}
where $\Delta _{f}=E_f-w_e$,  $\Delta _{kd}=E_{kd}-w_e$. We consider the emitter to be resonant with the flat band ($\Delta_f = 0$) as well as with the dispersive band at the $k_r$ mode. Eq.~(\ref{eq:eff}) is simplified as
\begin{equation}
H_{\mathrm{int}}=\frac{g}{\sqrt{2N}}\sum_k{\left( \sigma _-C_{k_f}^{\dagger}+\sigma _-C_{k_d}^{\dagger}e^{i\Delta _{kd}t} \right)}+\mathrm{H}.\mathrm{c}.
\end{equation}
In the single-excitation subspace, the state of the entire system is represented as 
\begin{equation}
	|\psi \left( t \right) \rangle =\sum_k{\left[ c_{k_f}(t)|g,1_{k_f}\rangle +c_{k_d}\left( t \right) |g,1_{k_d}\rangle \right]}+c_e\left( t \right) |e,0\rangle. \nonumber
\end{equation}
The initial state is set as $|e,0\rangle$, where the emitter is in the excited state and the lattice is in the vacuum state, i.e., $c_e(0)=1$. The evolution of the whole system governed by $H_{\mathrm{int}}$ is derived by solving the Schrödinger equation, i.e.,
\begin{gather}
\dot{c}_e\left( t \right) =-i\frac{g}{\sqrt{2N}}\sum_k{\left[ c_{k_f}\left( t \right) +e^{-i\Delta _{kd}t}c_{k_d}\left( t \right) \right]},\label{eq:e}
\\
\dot{c}_{k_f}\left( t \right) =-i\frac{g}{\sqrt{2N}}c_e\left( t \right),\label{eq:f}
\\
\dot{c}_{k_d}\left( t \right) =-i\frac{g}{\sqrt{2N}}e^{i\Delta _{kd}t}c_e\left( t \right).\label{eq:d}
\end{gather}

By substituting the integral form of Eqs.~(\ref{eq:f}) and (\ref{eq:d}) into Eq.~(\ref{eq:e}) and replacing $\sum_k$ with $\frac{N}{2\pi}\int dk$, the evolution of ${c}_e\left( t \right)$ is derived as 
\begin{align}
\dot{c}_e\left( t \right) &= -\frac{g^2}{2}\int_0^t{c_e\left( t^\prime \right) dt^\prime} \notag \\
	&\quad  -\frac{g^2}{2N}\frac{N}{2\pi}\int_{-\pi}^{\pi}{e^{i\varDelta _{kd}\left( t^\prime-t \right)}dk}\int_0^t{c_e\left( t^\prime \right) dt^\prime}. \label{eq:et}
\end{align}

We approximate the dispersion relation around ${\pm}k_r$ to be linear, that is,
\begin{equation}
	v_g=\frac{dE_d\left( k \right)}{dk} \bigg|_{k_r}=-4J\sin \left( k_r \right), 
\end{equation}
where $v_g$ is the group velocity at $k_r$. By setting $\delta _k=k-k_r$, the detuning is written as $\Delta _{kd}\simeq v_g\delta _k$. With Born-Markovian approximation, we extend the integral bound $\pm \pi$ to infinity. Consequently, Eq.~(\ref{eq:et}) is reduced to
\begin{equation}
\dot{c}_e\left( t \right) =-\frac{g^2}{2} \int_0^t{c_e\left( t^{\prime} \right) dt^{\prime}}-\frac{g^2}{2v_g} c_e\left( t \right) . 
\end{equation}
Taking the derivative of the formula, we derive
\begin{equation}
	\ddot{c}_e\left( t \right) +\frac{g^2}{2v_g}\dot{c}_e\left( t \right) +\frac{g^2}{2}c_e\left( t \right) =0.
\end{equation} 
By setting the initial condition $c_e\left( 0 \right) =1$ and $c_{k_f}(0)=c_{k_d}(0)=0$, we obtain $\dot{c}_e(0)=0$ from Eq.~(\ref{eq:e}) and
\begin{equation}
	c_e\left( t \right)=e^{-\frac{\Gamma }{2}t} \cos \left( \frac{\Omega}{2} t \right),\quad\Gamma  =\frac{g^2}{2v_g},\quad
	\Omega =\sqrt{2}g,\label{eq:ce(t)}
\end{equation}
where $\Gamma $ and $\Omega$ denote the decay rate and the Rabi oscillation rate, respectively. The decay rate $\Gamma$ is half of the decay rate for a small atom spontaneously dissipating into a vacuum bath. This reduction results from the transformation, where $a_x$ becomes $(C_{x,f}+C_{x,d})/\sqrt{2}$. The emitter interacts with the superposition state $(C_{x,f}+C_{x,d})$, causing the coupling strength to decrease to  $g/\sqrt{2}$. Eq.~(\ref{eq:ce(t)}) shows that the emitter evolves as a combination of Rabi oscillations and spontaneous decay. Part of the emitter's energy enters the flat band through the coupling point $C_{0,f}$, undergoing Rabi oscillations, i.e.,  $\mathrm{cos}(\Omega t)$. The remaining energy decays into the bath via the coupling point $C_{0,d}$. 

\begin{figure}[ht]
	\centering \includegraphics[width=\columnwidth]{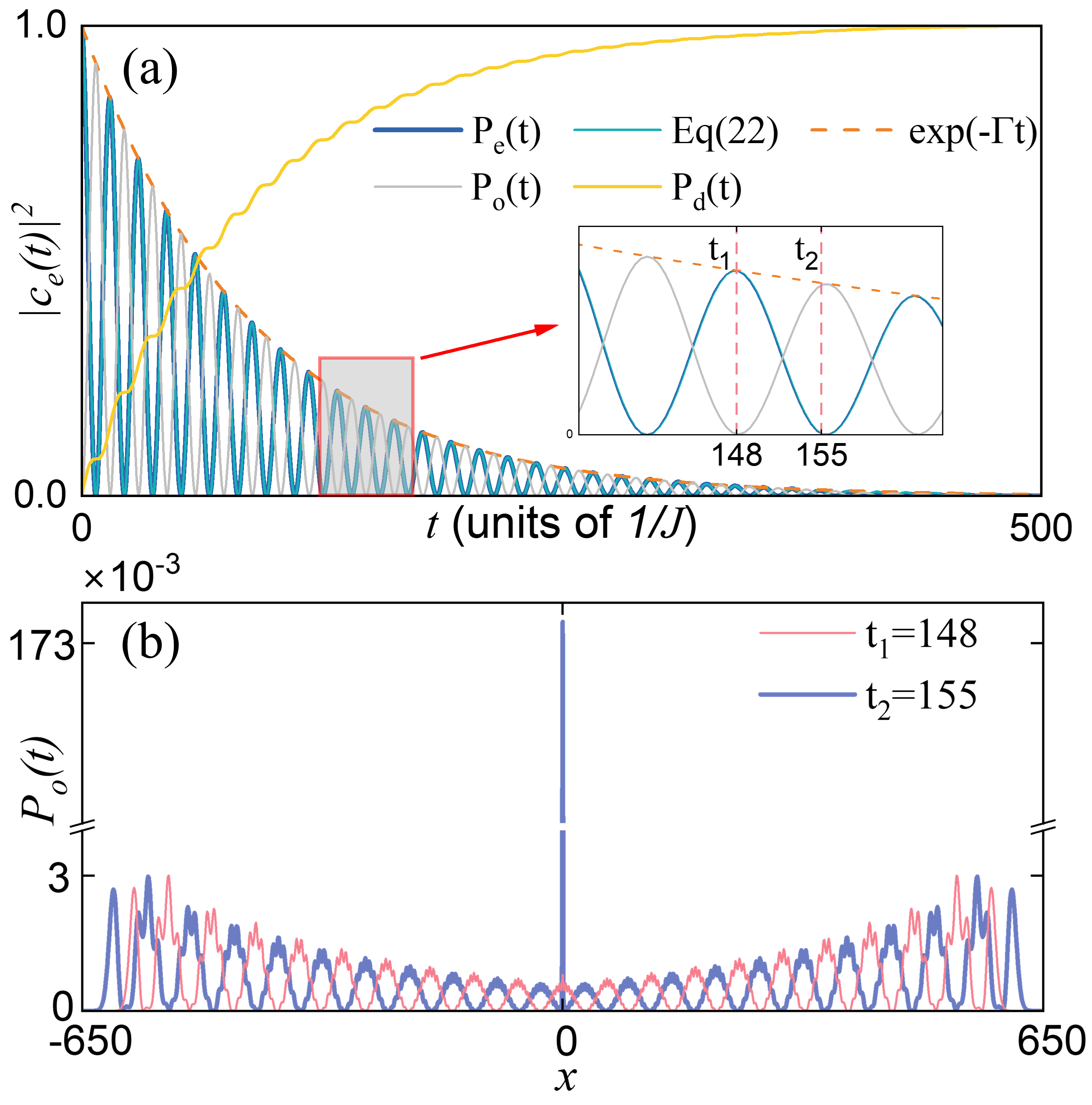}
	\caption{(a) Numerical and analytical results for the state population of a small emitter coupled to the flat band and the center of the dispersive band. The inset shows the evolution within the time interval $t_1=148$ to $t_2=155$. (b) Field amplitude at $t_1=148$ and $t_2=155$. The parameters used are $J=1$, $t=0$, $\omega_e=0$, and $g=0.3$.}
	
	\label{fig2}
\end{figure}

As shown in Fig.~\ref{fig2}(a), we present the numerical results for $P_e\left( t \right) =|c_e\left( t \right) |^2$, $P_0\left( t \right) =|c_{kf}\left( t \right) |^2=|c_{0,a}|^2+|c_{0,b}|^2
$ and $P_d\left( t \right) =\sum_k{|c_{kd}\left( t \right) |^2=}\sum_{n\ne 0}{|c_{n,a}|^2+|c_{n,b}|^2}$, alongside the analytical result from Eq.~(\ref{eq:ce(t)}). The numerical evolution of  $P_e\left( t \right) $ matches well with the analytical result. Over time, the atom's energy decreases, and the amplitude of the Rabi oscillations gradually diminishes until all the energy transfers into the lattice. The edges of the numerical curves are consistent with the analytical exponential decay $\mathrm{exp}(-\Gamma t)$. 

We select two specific time points, the trough $t_1=148$ and the peak $t_2=155$, and plot the field amplitudes in Fig.~\ref{fig2}(b). At $t_1$, most of the energy has oscillated back to the atom, so the amplitude at the coupling point is solely due to the decay contribution. In contrast, at $t_2$, the atom's energy undergoes Rabi oscillations with $C_f$, resulting in a significantly higher amplitude  $P_0(t)$. As time progresses, the energy transmits to further points, and the radiated wavepacket takes an exponential form modulated by $\mathrm{cos}(\Omega t)$. The simulation methods can be found in Appendix \ref{AppendixA}.
\begin{figure*}[ht]
	\centering \includegraphics[width=\textwidth]{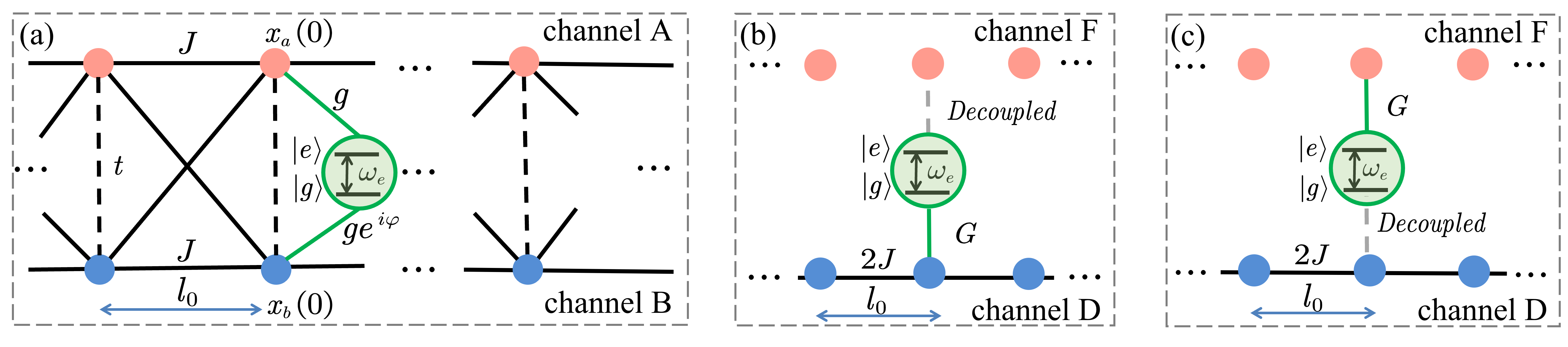}
	\caption{(a) A giant emitter interacts with the 1D stitch model at two coupling points, $x_a(0)$ and $x_b(0)$, with coupling strengths $g$ and $ge^{i\phi}$, respectively. (b) The equivalent model where the giant atom interacts exclusively with the dispersive band, decoupling from the flat band. The equivalent coupling strength is $G$. (c) The equivalent model where the giant atom interacts exclusively with the flat band, decoupling from the dispersive band.}
	
	\label{fig3}
\end{figure*}
\subsection{ Phase-Tuned Selective Interaction for Giant Atoms}
In the case of a single small atom, we find that it is equivalent to a giant atom in the equivalent lattice space, evolving through a combination of spontaneous decay and Rabi oscillation. While small atoms interact uniformly with all available modes, giant atoms introduce an additional degree of control via the relative phase between coupling points. This distinctive property enables selective interactions with specific bands, as demonstrated below.

As shown in Fig.~\ref{fig3}(a), we consider a giant emitter interacting with the lattice at two sites, $x_a(0)$ and $x_b(0)$, with a local phase $\phi$ encoded in the coupling. The interaction Hamiltonian in this configuration is given by
\begin{equation}
H_{\mathrm{int}}=g\sigma _-a_{0}^{\dagger}+ge^{i\phi}\sigma _-b_{0}^{\dagger}+\mathrm{H}.\mathrm{c}.\label{giantint} 
\end{equation}
Following a similar procedure, we use the transformation relationships to rewrite Eq.~(\ref{giantint}) in its equivalent form
\begin{equation}
H_{\mathrm{int}}=g\sigma _-\frac{\left( C_{0,f}^{\dagger}+C_{0,d}^{\dagger} \right)}{\sqrt{2}}+ge^{i\phi}\sigma _-\frac{\left( C_{0,d}^{\dagger}-C_{0,f}^{\dagger} \right)}{\sqrt{2}}+\mathrm{H}.\mathrm{c}. 
\end{equation}
By modulating the phase, the giant emitter can selectively interact with either the flat band or the dispersive band.

\subsubsection{Selective Interaction with Dispersive Band}	
First, we set the phase $\phi=0$. After applying the Fourier transform, the equivalent interaction Hamiltonian is expressed as
\begin{equation}
	H_{\mathrm{int}}= \frac{\sqrt{2}g}{\sqrt{N}}\sum_k{\left[ \sigma _-C_{_{kd}}^{\dagger}+\sigma _+C_{kd} \right]}.
\end{equation}
In this case, the intriguing phenomenon is that the influence of the flat band completely disappears, as shown in Fig.~\ref{fig3}(b). The giant emitter interacts only with the dispersive band, and the equivalent coupling strength becomes $\sqrt{2}g$. We then set $G=\sqrt{2}g$ and solve the Schrödinger equation to obtain the result for $c_e(t)$ in the interaction picture
\begin{gather}
\dot{c}_e\left( t \right) =-i\frac{G}{\sqrt{N}}\sum_k{e^{-i\Delta _{kd}t}c_{kd}\left( t \right)},\label{giant1}
\\
\dot{c}_{kd}\left( t \right) =-i\frac{G}{\sqrt{N}}e^{i\Delta _{kd}t}c_e\left( t \right), \label{giant2}
\\
\dot{c}_{kf}\left( t \right) =0.\label{giant3}
\end{gather}
We integrate Eqs.~(\ref{giant1}) and (\ref{giant2}), and substitute them into Eq.~(\ref{giant3}). The evolution equation becomes
\begin{equation}
	\dot{c}_e\left( t \right) =-\frac{G^2}{N}\sum_k{ \int_0^t{e^{-i\Delta _{kd}\left( t-t^\prime \right)}c_e\left( t^\prime \right) dt^\prime} }.
\end{equation}
By replacing $\sum_k$ with $\frac{N}{2\pi}\int dk$, we obtain
\begin{equation}
\dot{c}_e\left( t \right) =-\frac{G^2}{N}\frac{N}{2\pi}\int_{-\pi}^{\pi}{dk}\int_0^t{e^{i\varDelta _{kd}\left( t^{\prime}-t \right)}c_e\left( t^{\prime} \right) dt^{\prime}}.
\end{equation}
Using the Weisskopf-Wigner approximation, we reach
\begin{equation}
c_e\left( t \right) =\exp(-\frac{\Gamma}{2}t),\quad \Gamma =\frac{2G^2}{v_g}=\frac{4g^2}{v_g},\label{giant4}
\end{equation}
where $\Gamma$ is the spontaneous decay rate of the giant emitter, which is double that of the small emitter coupled to a 1D ladder lattice. This is because the coupling strength with the dispersive band is larger, given by $\mathrm{G}=\sqrt{2}g$. In Fig.~\ref{fig4}(a), we plot both the analytical evolution from Eq.~(\ref{giant4}) and the numerical evolution. We find that they match well, with the giant atom interacting only with the dispersive band and experiencing spontaneous decay.

\subsubsection{Selective Interaction with Flat Band}
\begin{figure}[ht]
	\centering 
	\includegraphics[width=\columnwidth]{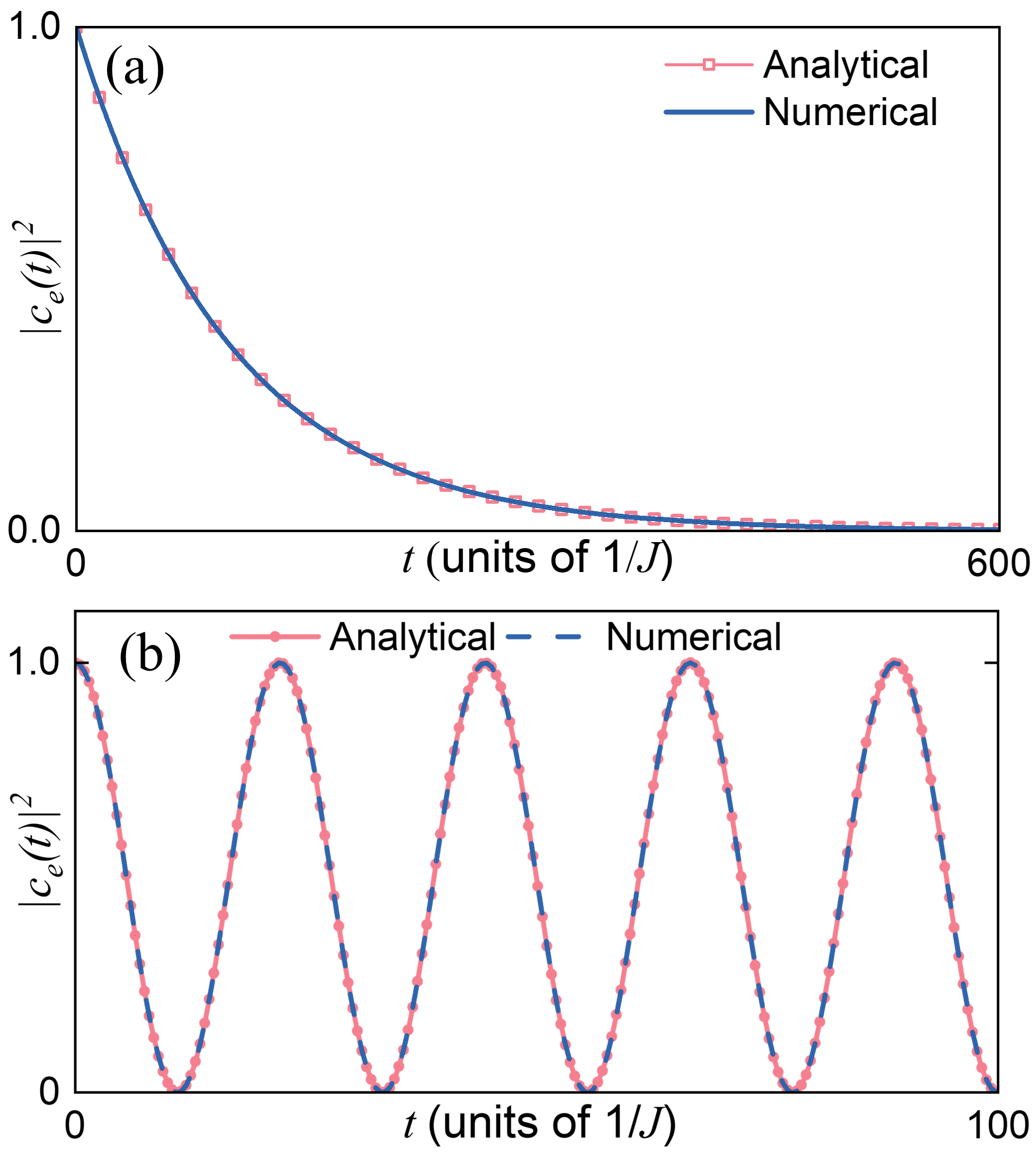}
	\caption{The evolution of $P_e(t)$ for a giant emitter inteacting with the flat band when $\phi=0$ (a), and  the dispersive band when $\phi=\pi$ (b). Parameters of the whole system are $g=0.1$, $J=1$, and $t=0$.}
	\label{fig4}
\end{figure}

In this section, we set the phase to $\pi$. Similar to the previous section, the equivalent interaction Hamiltonian becomes
\begin{equation}
	H_{\mathrm{int}}= \frac{G}{\sqrt{N}}\sum_k{\left[ \sigma _-C_{kf}^{\dagger}+\sigma _+C_{kf} \right]}.\label{giant5}
\end{equation}
The interaction with the dispersive band disappears during the process, and the giant emitter interacts exclusively with the flat band, as depicted in Fig.~\ref{fig3}(c). The evolution of the excited-state population, $P_e\left( t \right)=|c_e\left( t \right) |^2$, is obtained by solving the Schrödinger equation in the interaction picture. Substituting Eq.~(\ref{transfor mattrix}) into Eq.~(\ref{giant5}) and rewriting the interaction Hamiltonian in the interaction picture, we arrive
\begin{equation}
H_{\mathrm{int}}=\frac{G}{\sqrt{N}}\sum_k{\left( \sigma _-C_{kf}^{\dagger}e^{i\Delta _{f}t}+\sigma _+C_{kf}e^{-i\Delta _{f}t} \right)},
\end{equation}
where $\Delta _{f}=E_f -\omega _e$ is the detunnig from the flat band. Substituting the Hamiltonian into the Schrödinger equation, we obtain
\begin{gather}
	\dot{c}_e\left( t \right) =-i\frac{G}{\sqrt{N}}\sum_k{e^{-i\Delta _{f}t}c_{kf}\left( t \right)},\label{giantf1}\\
	\dot{c}_{kf}\left( t \right) =-i\frac{G}{\sqrt{N}}e^{i\Delta _{f}t}c_e\left( t \right),\label{giantf2} \\
	\dot{c}_{kd}\left( t \right) =0\label{Eq35}.
\end{gather}
By defining $$	\tilde{c}_e\left( t \right) =e^{i\frac{\Delta _{f}}{2}t}c_e\left( t \right),\quad \tilde{c}_{kf}\left( t \right) =e^{-i\frac{\Delta _{f}}{2}t}c_{kf}\left( t \right), $$  Eqs.~(\ref{giantf1}) - (\ref{giantf2}) become
\begin{gather}
	\frac{d\tilde{c}_e\left( t \right)}{dt}=i\frac{\Delta _{f}}{2}\tilde{c}_e\left( t \right) -i\frac{G}{\sqrt{N}}\sum_k{\tilde{c}_{kf}\left( t \right)},\label{giantf3}
	\\
	\frac{d\tilde{c}_{kf}\left( t \right)}{dt}=-i\frac{\Delta _{f}}{2}\tilde{c}_{kf}\left( t \right) -i\frac{G}{\sqrt{N}}\tilde{c}_e\left( t \right) .\label{giantf4}
\end{gather}
Eq.~(\ref{giantf3}) - Eq.~(\ref{giantf4}) result in
\begin{gather}
	c_e\left( t \right) =e^{i\frac{\Delta _{f}}{2}}\left[ \cos \left( \frac{\tilde{\Omega}_n}{2}t \right) -i\frac{\Delta _{f}}{\tilde{\Omega}_n}\sin \left( \frac{\tilde{\Omega}_n}{2}t \right) \right],\nonumber\\
	 \tilde{\Omega}_n=\sqrt{\Delta_f ^2+(2G)^2}.\label{giantf5}
\end{gather}
In Fig.~\ref{fig4}(b), we plot the analytical evolution described by Eq.~(\ref{giantf5}) alongside the numerical results, which match well with each other. 

\section{Emitter Frequency in the Band Gap: Bound States and Dipole-Dipole Interaction}
\begin{figure*}[ht]
	\centering \includegraphics[width=\textwidth]{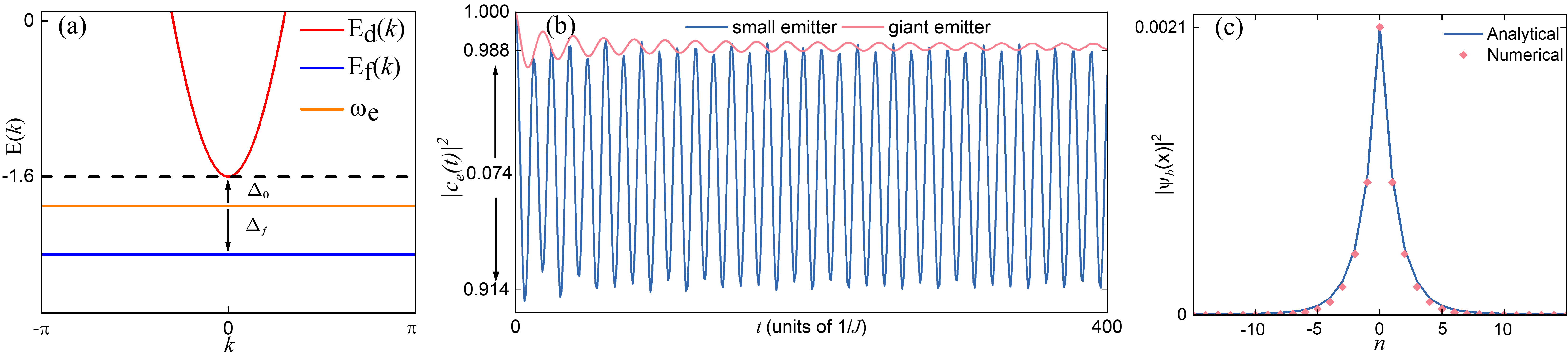}
	\caption{(a) The energy bands of the 1D ladder lattice and the emitter's frequency. The emitter's frequency is located in the band gap and near the lower edge of the dispersive band. (b) Numerical results for the excited-state population $P_e(t)=|c_e(t)|^2$ of the small and giant emitters as a function of time $t$. For sufficiently large $t$, $P_e(t)$ of the small emitter oscillates around a steady value, fitting well with the bound state of the giant emitter. (c) The field distributions of sublattices $x_a(n)$ and $x_b(n)$ when the giant emitter couples to sites $x_a(0)$ and $x_b(0)$, forming a bound state. The parameters for the small emitter are $\omega_e=-1.9$, $t=2.4$, $J=1$, and $g=0.1$. For the giant emitter, $\phi=0$ and $g=0.05$, with other parameters unchanged.
	\label{fig5}
	}
	
\end{figure*}

\subsection{Single atom in bound states}
A band gap appears when $t<-2J$. We now set the emitter's frequency $\omega_e$ within the gap. As illustrated in Fig.~\ref{fig5}(a), $\omega_e$ is close to the lower bound of the dispersive band, and the frequency detuning $\Delta_0$  is significantly smaller than $\Delta_f$.  
For a small emitter, we adopt the approximation that the emitter interacts independently with the flat and dispersive bands. The resulting dynamics are equivalently the product of two independent processes. In this scenario, we employ the same approach to analyze the emitter's dynamics. The contribution from the flat band results in a detuned Rabi oscillation, and the evolution of the excited-state population is given by
\begin{gather}
     |c_e\left( t \right) |^2=1-A\sin ^2\left( \frac{\Omega _f}{2}t \right),\nonumber\\ A=\frac{2g^2}{2g^2+\Delta _{f}^{2}},\label{bound0}\\
	 \Omega _f=\sqrt{2g^2+\Delta _{f}^{2}}.\label{smallwf}
\end{gather} 
The oscillation amplitude $A$ is determined by the detuning parameter $\Delta_f$, while the Rabi frequency $\Omega_f$, characterizing the interaction rate between the atoms and the flat band, also depends on $\Delta_f$.
The detuning from the dispersive band leads to the formation of bound states for the emitter. In this scenario, the interaction Hamiltonian includes only the contribution from the dispersive band, given by 
\begin{equation}
	H_{\mathrm{int}}=\mathcal{G} \sum_k{\left( \sigma _-c_{k_d}^{\dagger}+\sigma _+c_{k_d} \right)},\quad 
\end{equation}
where we set $\mathcal{G}  =g/\sqrt{2N}$ for the small atom and $\mathcal{G}  =\sqrt{2}g/\sqrt{N}$ for the giant atom. Similar to the derivations for Eqs.~(\ref{giantf1}) - (\ref{Eq35}), we obtain differential equations for $c_e(t)$ and $c_{kf(d)}(t)$. Defining $\tilde{C}_{kf\left( d \right)}\left( t \right) =c_{kf\left( d \right)}\left( t \right) \exp(-i\Delta _{kd}t)
$, the evolution is derived in Laplace space with $c_e\left( s \right) =\int_0^{\infty}{c_e\left( t \right) e^{-st}dt}$, $\tilde{C}_{kf\left( d \right)}\left( s \right) =\int_0^{\infty}{\tilde{C}_{kf\left( d \right)}\left( t \right) e^{-st}dt}
$, we obtain
\begin{gather}
s\tilde{c}_e\left( s \right) -c_e\left( 0 \right) =-i\sum_k{ \mathcal{G} \tilde{C}_{kd}\left( s \right) \,\,}
\\
s\tilde{C}_{kd}\left( s \right) -C_{kd}\left( 0 \right) =-i\Delta _{kd}\tilde{C}_{kd}\left( s \right) -i \mathcal{G} \tilde{c}_e\left( s \right) \label{bound}
.
\end{gather}
Using the primary condition $c_e\left( 0 \right) =1$ and $c_{kd}\left( 0 \right) =0$, Eq.~(\ref{bound}) is
\begin{equation}
\tilde{C}_{kd}\left( s \right) =\frac{-i \mathcal{G} \tilde{c}_e\left( s \right)}{s+i\Delta _{kd}},
\end{equation}
and
\begin{gather}
\tilde{c}_e\left( s \right) =\frac{1}{s+\varSigma _e\left( s \right)},\\
\Sigma _e\left( s \right) =\sum_k{\frac{ \mathcal{G} ^2}{s+i\Delta _{kd}}},
\end{gather}
where $\Sigma _e\left( s \right)$ is the self-energy. The time-independent evolution can be obtained via the inverse Laplace transform
\begin{equation}
c_e\left( t \right) =\frac{1}{2\pi i}\underset{E\rightarrow \infty}{\lim}\int_{\varepsilon -iE}^{\varepsilon +iE}{\tilde{c}_e\left( s \right) e^{st}ds},\quad \varepsilon >0.
\end{equation}
Assuming the ladder is sufficiently long that the emitted field cannot touch the open boundary condition within the considered time, we can rewrite the self-energy in integral form by replacing $\sum_k$ with $\frac{N}{2\pi}\int dk$. Under this substitution, the self-energy is expressed as
\begin{equation}
\Sigma _e\left( s \right) =\frac{N}{2\pi}\int_{-\pi}^{\pi}{\frac{ \mathcal{G} ^2}{s+i\Delta _{kd}}dk},\label{bound1}
\end{equation}
Around $k_{\mathrm{min}}$, the dispersion relation can be approximated as a quadratic form
\begin{equation} 
	E_d\left( k \right)=E_d\left( k_{\mathrm{min}} \right) + \alpha (k - k_{\mathrm{min}})^2. \label{bound2}
\end{equation}
The curvature $\alpha$ at $k_{\mathrm{min}}$ is defined as the second derivative of $E_d(k)$, expressed as
\begin{equation} \alpha = \frac{1}{2} \left. \frac{d^2E_d\left( k \right)}{dk^2} \right|_{k=k_{\mathrm{min}}} = 2J.
\end{equation}
By substituting Eq.~(\ref{bound2}) into Eq.~(\ref{bound1}), the self-energy is calculated as
\begin{equation}
\Sigma _e\left( s \right) =\frac{N \mathcal{G} ^2}{2\pi}\int_{-\infty}^{\infty}{\frac{1}{s+i\left[ \Delta _0+\alpha k^2 \right]}dk},
\end{equation}
where $\Delta _0=E_d\left( k \right) _{\min}-\omega _q$ is the detuning from the dispersive band edge. We assume that $\Delta _0$ is small, and only the modes around $k_{\min}$ are excited with high probabilities. Consequently, we obtain 
\begin{equation}
	\Sigma _e\left( s \right) =\frac{N \mathcal{G} ^2}{2\sqrt{-\alpha \left( \Delta _0-is \right)}}.
\end{equation}
Using the residue theorem, the steady-state probability is calculated~\cite{Calajo2016}
\begin{gather}
	|c_e\left( t=\infty \right) |^2=|\mathrm{Re}s\left( s_0 \right) |^2,\label{bound3}
	\\
	\mathrm{Re}s\left( s_0 \right) =\left. \frac{1}{1+\partial _s\Sigma _e\left( s \right)} \right|_{s=s_0},
\end{gather}
where $\mathrm{Re}s\left( s_0 \right)$ denotes the steady-state population for the small atom, and $s_0$ is the purely imaginary pole of the transcendental equation, which is determined by solving~\cite{Wang_2022chiral}
\begin{equation}
	s_0+\Sigma _e\left( s_0 \right) =0.
\end{equation}

In Fig.~\ref{fig5}(b), we show the numerical evolution of $P_e\left( t \right)=|c_e\left( t \right) |^2$ for both a giant atom and a small atom. For sufficiently large $t$, $P_e\left( t \right)$ of the small atom reaches a steady value of approximately 0.988, which corresponds to the bound state. This value is in good agreement with the analytical result from Eq.~(\ref{bound3}), which predicts $P_e\left( t \right)=0.989$. However, in contrast to conventional bound states in traditional models, $P_e\left( t \right)$ oscillates around this steady value due to the interaction with the flat band. The amplitude of these oscillations, as determined by Eq.~(\ref{bound0}), is about 0.074, which is consistent with the numerical result, as shown in the figure. To further validate our approximation, we present the numerical result for the bound state of the giant emitter, where the frequency of the giant emitter is set to match that of the small emitter, and it interacts solely with the dispersive band. From these numerical results, we observe that the steady-state value of the small atom matches well with that of the giant emitter bound state.

The field transport is described by
\begin{equation}
\psi _b\left( x \right) \propto \exp \left( -\frac{x}{L_{\mathrm{eff}}} \right),
\end{equation}
where $L_{\mathrm{eff}}=\sqrt{\frac{\alpha}{\Delta _0}}$ is the effective propagation length of the bound state within the sublattice. In Fig.~\ref{fig5}(c), we compare the numerical results with the analytical expression, showing excellent agreement.

\subsection{Dipole-dipole interactions}
\begin{figure*}[ht]
	\centering
	\includegraphics[width=\textwidth]{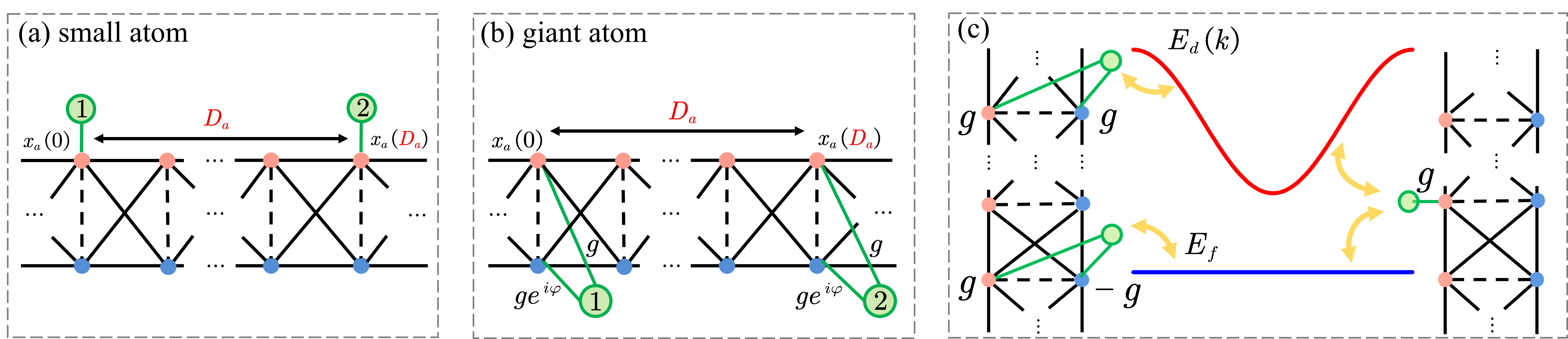}
	\caption{(a) Two small and (b) two giant atoms, which are separated by distance ${\color[RGB]{240, 0, 0} D_a}$, interact with the 1D cross-stitch lattice. For giant atoms, the coupling points are located at different sublattice $A$ and $B$. (c) Schematic illustration of the interaction between a small atom or a giant atom and the energy bands. The giant atom enables selective interaction: for $\phi=0$, it couples exclusively to the dispersive band, while for $\phi=\pi$ it interacts solely with the flat band. In contrast, the small atom exhibits non-selective coupling, interacting simultaneously with both the dispersive band and the flat band. 
		\label{fig6}
	}
\end{figure*}
Due to the exponential localization, the bound states of two atoms overlap when their separation distance $D_a$ is sufficiently small, leading to dipole-dipole interactions. As illustrated in Fig.~\ref{fig6}(a), we consider two atoms located within the same bandgap region. They interact with the cross-stitch lattice and are separated by a distance $D_a$. Similar to the previous discussion, we focus on the differences between the giant and small atom configurations.
\subsubsection{Dipole-dipole interactions between two small atoms}
\begin{figure}[ht]
	\centering \includegraphics[width=\columnwidth]{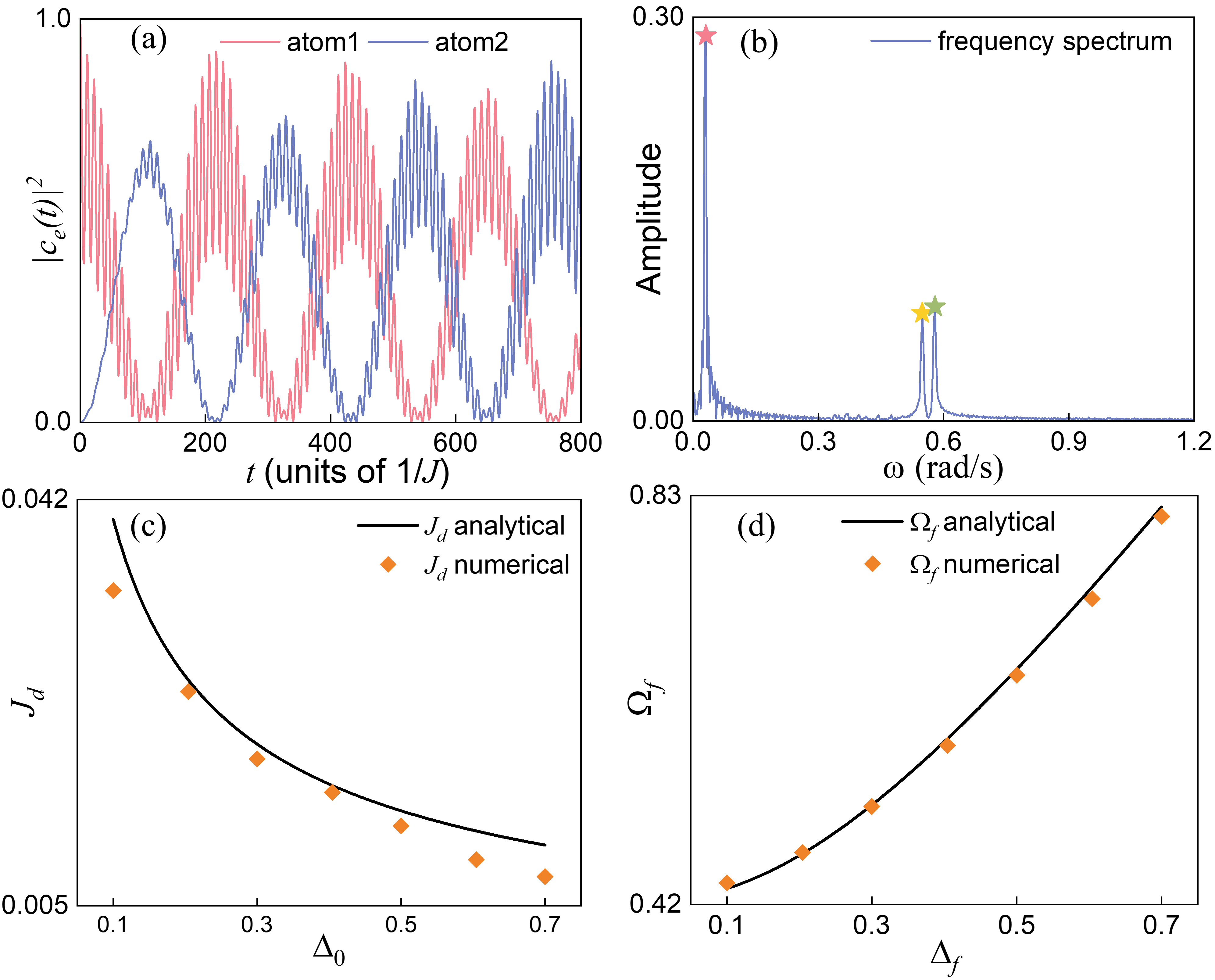}
	\caption{(a) The dynamics of the two small atoms. (b) The frequency components of atomic dynamical evolution via Fourier transformation. Here we set $\omega_e=-2$. (c) The dipole-dipole interaction strength between two small atoms as a function of detuning $\Delta_0$. (d) The Rabi frequency as a function of detuning $\Delta_f$. The parameters are $g = 0.3$, $D_a = 1$, and others are consistent with Fig.~\ref{fig5}.
		\label{fig7}
	}
	
\end{figure}

Both small atoms are assumed to couple to sublattice A at two distinct sites, separated by a distance $D_a$. Similar to the single-atom case, we adopt the assumption that the small atoms interact independently with the flat and dispersive bands while exchanging virtual photons through the dispersive band. Under this assumption, the interaction Hamiltonian is expressed as
\begin{gather} H_{\mathrm{int},2}=\frac{g_{\mathrm{eff},i}}{\sqrt{N}}\sum_i{\sum_k{\left( \sigma_i^{-}C_{kd}^{\dagger} + \sigma_i^{-}C_{kf}^{\dagger} \right)}} + \mathrm{H}.\mathrm{c.}, \nonumber \\  g_{\mathrm{eff},i}(k)=g_{\mathrm{eff}}e^{-ikx_i},
\end{gather}
where the first term contributes to the effective interaction between the two small atoms, while the second term results in Rabi oscillations with the flat band. Here, we primarily focus on the dipole-dipole interaction mediated by the dispersive band modes, and therefore, neglect the second term. In the rotating frame, the interaction Hamiltonian becomes
\begin{equation}
H_{\mathrm{int},2}\left( t \right) =\frac{g_{\mathrm{eff},i}}{\sqrt{N}}\sum_i{\sum_k{\sigma _{i}^{-}C_{kd}^{\dagger}e^{i\Delta _{kd}t}}}+\mathrm{H}.\mathrm{c}., 
\end{equation}
where $\Delta _{kd}=E_d\left( k \right) -\omega _e$. By employing the effective Hamiltonian methods~\cite{james2007effective}, the one-mode-mediated effective Hamiltonian can be expressed as
\begin{align}
	H_{\mathrm{eff}} = \sum_k & \frac{g_{\mathrm{eff},1}g_{\mathrm{eff},2}^{*}}{\Delta _{kd}N}
	\left( \sigma _{1}^{-}C_{kd}^{\dagger}\sigma _{2}^{\dagger}C_{kd} \right. \notag \\
	&\left. - \sigma _{2}^{\dagger}C_{kd}\sigma _{1}^{-}C_{kd}^{\dagger} \right)
	+ \mathrm{H}.\mathrm{c}.
\end{align}

At $t=0$, one atom is in the excited state while the other remains in the ground state. There exist a Rabi oscillation between the two atoms, while the lattice remains virtually excited and approximately in the vacuum state. Hence, we adopt the approximation
\begin{equation}
	\langle C_{kd}^{\dagger}C_{kd}\rangle \simeq 0,\quad\langle C_{kd}C_{kd}^{\dagger}\rangle \simeq 1.
\end{equation}
Then, $H_{\mathrm{eff}}$ is simplified as
\begin{equation}
H_{\mathrm{eff}}\simeq \sum_k{\frac{g_{\mathrm{eff},1}g_{\mathrm{eff},2}^{*}}{\Delta _{kd}N}\sigma _{2}^{\dagger}\sigma _{1}^{-}}+\mathrm{H}.\mathrm{c}.
\end{equation}
We obtain the interaction strength 
\begin{equation}
	J_{d}=-\sum_k{\frac{g_{\mathrm{eff},1}g_{\mathrm{eff},2}^{*}}{\Delta _{kd}N}}\simeq -\frac{1}{2\pi}\int_{-\pi}^{\pi}{\frac{g_{\mathrm{eff},1}g_{\mathrm{eff},2}^{*}}{\Delta _{kd}}dk}.
\end{equation}
Since the emitter's frequency lies below the edge of the dispersive band, the dispersion relation can be approximated as a quadratic form in Eq.~(\ref{bound2}). Substituting this into the expression for $J_{d}$, we obtain
\begin{equation}
	J_{d}=-\frac{g_{\mathrm{eff}}^{2}}{2\pi}\int_{-\pi}^{\pi}{\frac{e^{-ik D_a}}{\Delta _0+\alpha k^2}dk}.
\end{equation}
Finally, the dipole-dipole interaction strength is derived as
\begin{equation}
	J_{d}=-\frac{g_{\mathrm{eff}}^{2}}{2\sqrt{\Delta _0\alpha}}e^{-\sqrt{\frac{\Delta _0}{\alpha}} D_a},\label{J_d}
\end{equation}
which indicates that $J_{d}$  is modulated by the
detuning $\Delta _0$ and the separation distance $D_a$ between the emitters. 

Assuming that atom 1 is initially excited, we plot the dynamics of the two atoms through numerical simulations in Fig.~\ref{fig7}(a). The results reveal that the two atoms leak energy into the lattice while exchanging photons. During this exchange process, both atoms independently undergo Rabi oscillations with the flat band. To further analyze the dynamics, we conduct a frequency-spectrum analysis on the evolution of the two atoms. The first peak in the spectrum corresponds to the dipole-dipole exchange frequency, $\omega_{d}=2J_{d}$, while the second and third peaks represent the frequencies of the Rabi oscillations.

Using the frequency-spectrum analysis, we extract the numerical frequencies of the dipole-dipole interaction and the Rabi oscillations. Fig.~\ref{fig7}(c) shows the numerical results for $J_{d}$ as a function of detuning $\Delta_0$, which are in good agreement with the analytical description given by Eq.~(\ref{J_d}). Fig.~\ref{fig7}(d) shows that the analytical Rabi oscillation frequencies (given by Eq.~(\ref{smallwf})) match well with the numerical results. These findings confirm that the two atoms interact with the flat band independently while exchanging virtual photons through the dispersive band. For small atoms, non-selective interaction leads to partial energy leakage into the flat band, thereby reducing the interaction fidelity. 
\subsubsection{Selective dipole-dipole interactions between two giant atoms}
\begin{figure*}[ht]
	\centering \includegraphics[width=\textwidth]{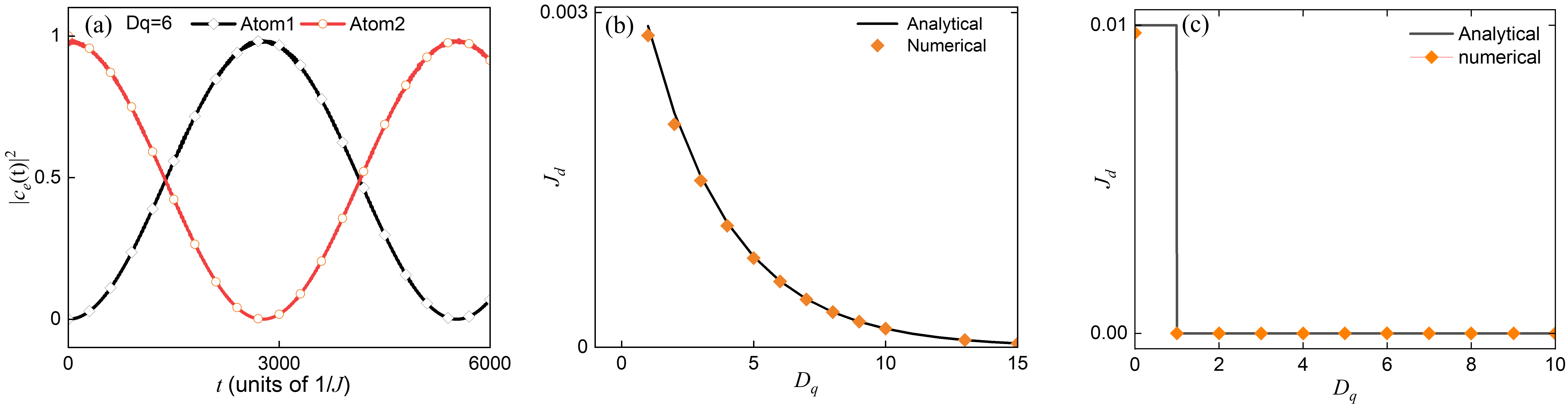}
	\caption{(a) Rabi oscillations between the two giant atoms. The parameters are $\omega_e = -1.8$, $D_a = 6$, and $g = 0.05$. (b) The Rabi frequency of the two giant emitters as a function of $D_a$. (c) Two giant atoms couple to the flat band. If $D_a > 0$, there is no dipole-dipole interaction between the two atoms. The parameters are $\omega_e = -1.9$, $g = 0.05$, with other parameters consistent with Fig.~\ref{fig5}.
		\label{fig8}
	}
\end{figure*}

We show that interference effects enable giant atoms to selectively interact with two distinct bands by modulating the coupling phase difference. Specifically, by setting $\phi = 0$, the giant atoms interact exclusively with the dispersive band, as illustrated in Fig.~\ref{fig6}(b). We now analyze the interaction between two giant atoms coupled to both sublattices A and B. The corresponding interaction Hamiltonian is 
\begin{gather} 
	H_{\mathrm{int},2} = \frac{G_{\mathrm{eff},i}}{\sqrt{N}} \sum_i \sum_k \sigma_{i}^{-} C_{kd}^{\dagger} + \mathrm{H.c.}, \nonumber \\ 
 G_{\mathrm{eff},i}(k)=G_{\mathrm{eff}}e^{-ikx_i}.
\end{gather}
Similar to the derivation for small atoms, the dipole-dipole interaction strength between two giant atoms is given by
\begin{equation}
	J_{d}=-\frac{G_{\mathrm{eff}}^{2}}{2\sqrt{\Delta _0\alpha}}e^{-\sqrt{\frac{\Delta _0}{\alpha}}D_a}.\label{giantJ_d}
\end{equation}
In Fig.~\ref{fig8}(a), we present the dynamics of two giant emitters obtained from numerical simulations. The results show that the two atoms coherently exchange excitations without decay. Furthermore, Fig.~\ref{fig8}(b) illustrates the variation of $J_{d}$ with the separation distance $D_a$, showing excellent agreement between the numerical results and the analytical expression in Eq.~(\ref{giantJ_d}). Unlike the case of small atoms, giant atoms can achieve high-fidelity energy exchange when tuned to a specific relative phase. 

When the relative phase is set to $\phi=\pi$, the two giant atoms couple exclusively to the flat band. The interaction Hamiltonian is
\begin{equation}
	H_{\mathrm{int},2}=\frac{G_{\mathrm{eff},i}}{\sqrt{N}}\sum_i{\sum_k{\sigma _{i}^{-}C_{kf}^{\dagger}}}+\mathrm{H}.\mathrm{c}. 
\end{equation}
Using the effective Hamiltonian theory, the dipole-dipole interaction strength between the two atoms is expressed as
\begin{equation}
	J_{d}=-\sum_k{\frac{G_{\mathrm{eff}}^{2}e^{-ikD_a}}{N\Delta _f}},
\end{equation}
where $\Delta_f = E_f(k) - \omega_e$ is a constant detuning. After simplification, $J_{d}$ becomes
\begin{equation}
	J_{d}=-\frac{G_{\mathrm{eff}}^{2}}{\Delta _f}\delta \left(D_a\right). 
\end{equation}
Since the eigenstates of the flat band are compact localized and only distributed at $x_a(n)$ and $x_b(n)$, the coupling strength $J_d$ vanishes when the giant atoms are not coupled to the same site. This result is consistent with previous studies, which demonstrate that when nearest-neighbor CLSs are orthogonal, photonic-mediated interactions remain strictly local and do not extend beyond adjacent unit cells \cite{Benedetto2024}. We plot the numerical and analytical results of $J_{d}$ as a function of the separation distance $D_a$ in Fig.~\ref{fig8}(c), demonstrating consistency between the two approaches.

\section{conclusion}
In this work, we explore the selective interaction capabilities of giant atoms in a 1D cross-stitch ladder lattice featuring a dispersive band and a flat band with tunable relative positions.
When the two bands intersect, small atoms interact with both bands simultaneously, while giant atoms selectively couple to either the dispersive or flat band, determined by the relative phase ($\phi=0$ or $\phi=\pi$) between their coupling points. This demonstrates the unique selective coupling of giant atoms.

For emitter frequencies within the bandgap, atom-photon bound states form. Small atoms, due to non-selective coupling, exhibit Rabi oscillations mediated by the flat band, leading to limited energy exchange fidelity. In contrast, giant atoms enable high-fidelity long-range interactions when $\phi=0$ by suppressing the flat band’s influence, or eliminate interactions entirely when $\phi=\pi$, providing precise control over interaction dynamics.

The relative phase $\phi$ of giant atoms can be tuned in superconducting quantum circuits by modulating the couplers connecting the atoms and sublattices~\cite{wang2022giant,roushan2017}. These findings underscore the versatility of giant atoms, where interference effects not only allow for flexible quantum control but also pave the way for designing selective quantum systems with potential applications in quantum information processing.

\section{Acknowledgments}
The quantum dynamical simulations are based on open source code 
QuTiP. 
X.W.~is supported by
the National Natural Science
Foundation of China (NSFC) (Grant No.~12174303 and No.~11804270), and China Postdoctoral Science Foundation (No.~2018M631136). 

\appendix
\section{The numerical simulation method}\label{AppendixA}
We simulate the interaction between a two-level emitter and the 1D cross-stitch lattice in real space. The numerical calculations proceed through the following steps:

(a) In our simulation, the Hilbert space is restricted within the single-excitation subspace, i.e., 
\begin{equation}
 |\psi \left( t \right) \rangle =c_e\left( t \right) |e,0\rangle +\sum_{x_i=1}^N{\sum_{i=a,b}{c_{x,i}\left( t \right) |g,1_{x_i}\rangle}}.\label{real-space}
\end{equation}
Here, $c_e\left( t \right)$ denotes the amplitude of the emitter in the excited state, and $c_{x,i}\left( t \right)$ ($i=a,b$) represents the amplitude of a single photon at $x_i$ in the sublattice A(B). The Hamiltonian in Eq.~(\ref{H_s}) can be expanded in the basis of Eq.~(\ref{real-space}). For a finite system with $N$ unit cells (corresponding to $2N$ lattice sites), the Hamiltonian can be represented as a matrix with dimensions $2N+Q$, with $Q$ as the number of atoms. To illustrate, we set $Q=1$ and $N=1500$, which is sufficiently large to prevent the propagating field from reaching the lattice boundary. Then, the Hamiltonian matrix is given by

\begin{widetext}
	\begin{equation}
 H_{\left( 2N+1,2N+1 \right)}=\left( \begin{array}{ccc}
	H_A&		H_{AB}&		0\\
	H_{BA}&		H_B&		0\\
	0&		0&		\omega _e\\
\end{array} \right) +\left( \begin{array}{ccccc}
	0&		0&		\cdots&		0&		0\\
	0&		0&		0&		0&		\vdots\\
	\vdots&		\vdots&		\ddots&		\vdots&		g\\
	0&		0&		0&		0&		\vdots\\
	0&		\cdots&		g&		\cdots&		0\\
\end{array} \right) ,
		\label{eq:full_matrix}
	\end{equation}

	\begin{equation}
	 H_A=H_B=\left( \begin{array}{ccccc}
				0&		-J&		0&		\cdots&		-J\\
				-J&		0&		-J&		\cdots&		0\\
				0&		-J&		0&		\cdots&		0\\
				\vdots&		\vdots&		\vdots&		\ddots&		\vdots\\
				-J&		0&		0&		-J&		0\\
			\end{array} \right) ,\quad H_{AB}=H_{BA}^{\dagger}=\left( \begin{array}{ccccc}
				-t&		-J&		0&		\cdots&		-J\\
				-J&		-t&		-J&		\cdots&		0\\
				0&		-J&		-t&		\cdots&		0\\
				\vdots&		\vdots&		\vdots&		\ddots&		\vdots\\
				-J&		0&		0&		-J&		-t\\
			\end{array} \right) ,
	\end{equation}
\end{widetext}
where $H_A$ ($H_B$) and $H_{AB}$ ($H_{BA}$) are $N\times N$ square matrices. The sub-diagonal terms of $H_A$ ($H_B$) represent inter-cell hopping within sublattice $A$ ($B$). The diagonal terms of $H_{AB}$ ($H_{BA}$) are intra-cell hopping between sublattices $A$ and $B$. The elements in the top right and bottom left corner of $H_A$ ($H_B$) and $H_{AB}$ ($H_{BA}$ are the periodic boundary conditions. The $g$ terms are the coupling strength between the atom and site $x_a(N/2)$. 

(b) Assuming that the atom (lattice) is initially in the excited (vacuum) state, i.e., $|\psi(t=0)\rangle=|e,0\rangle$, we employ the QuTiP package \cite{Johansson2012,Johansson2013} to numerically solve the time-dependent Schrödinger equation. This allows us to obtain the probability of the emitter $|c_e\left( t \right)|^2$,  as well as the photonic field distributions, $|c_{x,a}\left( t \right)|^2$ and $|c_{x,b}\left( t \right)|^2$.

Based on the method above, we plot all the dynamical evolutions in our work.

\section{The experimental setsups}
\begin{figure*}[ht]
	\centering \includegraphics[width=\textwidth]{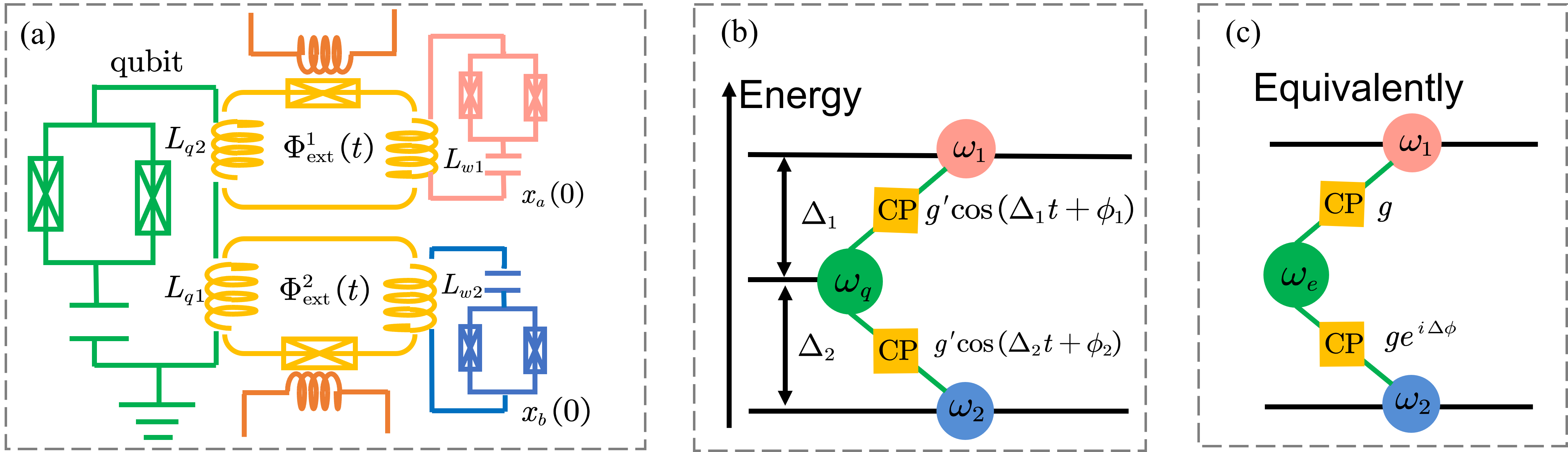}
	\caption{(a) Tunable coupling between two sublattice sites and a superconducting giant atom in the transmon form. The coupling points are mediated via loops with Josephson junctions (yellow crosses). At coupling point $i$, $L_{wi}$ and $L_{qi}$ represent the shared branch inductances of sublattice site $a(b)$ and giant atom, respectively. To realize time-dependent couplings, an external time-dependent flux $\Phi _{\mathrm{ext}}^{\left( i \right)} $ is applied. (b) A parametric modulation approach is employed to generate a tunable relative phase. Given that the frequency detuning between the atom and sublattice A (B) is $\Delta_1$ ($\Delta_2$), modulating the coupler that connects them at a frequency of $\Delta_1$ ($\Delta_2$) and phase $\phi_1$ ($\phi_2$) results in an effective resonance hopping. (c) The corresponding complex hopping amplitude for the cases $\Delta_1 = 0$ or $\Delta_2 = 0$ is shown, illustrating the effective resonant hopping.
		\label{fig9}
	}
	
\end{figure*}

In realistic experimental setups, arrays of transmons can be used to construct a 1D cross-stitch lattice waveguide discussed in this work~\cite{Mansikkam2022,Kim2021,Orell2019}. For giant atoms, the essence of selective interaction lies in the precise control of the relative phase $\phi$ between two coupling sites. As depicted in Fig.~\ref{fig9}(a), a superconducting giant atom interacts with two sublattice sites at positions $x_a(0)$ and $x_b(0)$. Each coupling point is mediated by a Josephon junction inserted in a loop. The inductances $L_{wi}$ and $L_{qi}$ of the $i$th loop ($i=1,2$) correspond to the shared branches in site $a(b)$ and the giant atom, respectively. The gauge-invariant phase difference across the Josephson junction in loop $i$ is denoted as $\phi _{J}^{\left( i \right)}$. The intermediate junction can be regarded as a lumped inductance $L_i$ as~\cite{Wulschner2016}
\begin{equation}
	 L_i=\frac{L_T}{\cos \phi _{J}^{\left( i \right)}},\quad L_T=\frac{\Phi _0}{2\pi I_c},
\end{equation}
where $\Phi _0=h/2e$ and $I_c$ denotes the critical current of the two junctions. The relation between $\phi _{J}$ and $\Phi _{\mathrm{ext}}^{\left( i \right)}$ follows from the expression for the total magnetic flux in the loop $\mathcal{C} _i$
\begin{equation}
 \phi _{J}^{\left( i \right)}=\int_{\mathcal{C} _i}{\mathbf{A}\mathrm{d}}\mathbf{l}=\frac{2\pi}{\Phi _0}\left[ \Phi _{\mathrm{ext}}^{\left( i \right)}-\left( L_{wi}+L_{qi} \right) I_c\sin \phi _{J}^{\left( i \right)} \right] ,
\end{equation}
this leads to
\begin{equation}
\phi _{J}^{\left( i \right)}+\beta \sin \phi _{J}^{\left( i \right)}=\frac{2\pi}{\Phi _0}\Phi _{\mathrm{ext}}^{\left( i \right)},\quad \beta =\frac{L_{wi}+L_{qi}}{L_T},
\end{equation}
which shows that $\phi _{J}$ can be controlled by the external flux. For simplicity, we assume $L_{wi}=L_{qi}=L_0$ and $\beta\ll 1$ (i.e., $L_0\ll L_T)$. Under this assumption, we obtain $\phi _{J}^{\left( i \right)}\simeq 2\pi\Phi _{\mathrm{ext}}^{\left( i \right)}/\Phi _0$. By applying the $Y - \Delta$ transformation to the coupling loop, the effective mutual inductance is derived as~\cite{Geller2015}
\begin{equation}
 M_{gi}=\frac{L_{0}^{2}}{2L_0+L_i}=\frac{L_{0}^{2}}{L_T}\cos \left( \frac{2\pi}{\Phi _0}\Phi _{\mathrm{ext}}^{\left( i \right)} \right),
\end{equation}
which shows that the mutual inductance $M_{gi}$ can be modulated by $\Phi _{\mathrm{ext}}^{\left( i \right)}$ in a cosine form where $\Phi _{\mathrm{ext}}^{\left( i \right)}$ is periodically modulated as
\begin{equation}
 \Phi _{\mathrm{ext}}^{\left( i \right)}=\Phi _{bi}+\frac{2\pi}{\Phi _0}d_i\cos \left( \Omega _{d}^{i}t+\phi _i \right) ,
\end{equation} 
where $\Phi _{bi}$ is the dc part, $d_i$ ($\phi_i$) is the modulating amplitude (phase) of the ac part at frequency $\Omega _{d}^{i}$. We analyze the frequency components of $M_{gi}(t)$ by expanding it in the Fourier form
\begin{equation}
	 M_{gi}\left( t \right) =\frac{L_{0}^{2}}{L_T}\sum_{n=0}^{\infty}{A_{i,n}\cos \left( n\Omega _{d}^{i}t+\phi _{i,n} \right)}.
\end{equation}
By numerically optimizing $\Phi _{bi}$, the dc component $A_{i,0}$, which represents the time-independent coupling inductance, can be eliminated. We assume that $d_i$ is small and neglect higher Fourier orders ($n\geqslant2$), considering only the fundamental frequency component $A_1$. Consequently, $M_{gi}$ can be approximated as~\cite{Wang_2022chiral}
\begin{equation}
M_{gi}\left( t \right) \simeq A_1\frac{L_{0}^{2}}{L_T}\cos \left( \Omega _{d}^{i}t+\phi _i \right) .
\end{equation}
According to the Josephson relation, the current operator of the transmon is approximately given by
\begin{equation}
 I_q=\sqrt{\frac{\hbar \omega _e}{2L_Q}}\left( \sigma _-+\sigma _+ \right) ,\quad 
\omega _e=\frac{1}{\sqrt{L_QC_q}}-\frac{E_C}{\hbar},
\end{equation}
where $L_Q$ is the total inductance of the transmon and $E_C=e^2/\left( 2C_q \right)$ represents the charging energy. The current operators of the quantized sites $a$ and $b$ are given by
\begin{equation}
I_{w1}=\sqrt{\frac{\hbar \omega _1}{2L_1}}\left( a+a^{\dagger} \right) ,\quad I_{w2}=\sqrt{\frac{\hbar \omega _2}{2L_2}}\left( b+b^{\dagger} \right) .
\end{equation}
$\omega_i$ and $L_i$ denote the on-site frequency and total inductance of site $i$ ($i=1,2$), respectively. Therefore, the interaction Hamiltonian is given by
\begin{align}
 H_I&=\sum_{i=1,2}{M_{gi}\left( t \right) I_qI_{wi}}\nonumber\\
	&\simeq \hbar \left[ g_1\left( t \right) \sigma _-a^{\dagger}+g_2\left( t \right) \sigma _-b^{\dagger}+\mathrm{H}.\mathrm{c}. \right] ,		
\end{align}
where the coupling strength $g_i\left( t \right)$ is 
\begin{align}
 g_i\left( t \right) & =\frac{1}{2}\sqrt{\frac{\omega _e\omega _i}{L_QL_i}}M_{gi}\left( t \right) =g_i\cos \left( \Omega _{d}^{i}t+\phi _i \right) ,\nonumber\\
g_i& =\frac{1}{2}\sqrt{\frac{\omega _e\omega _i}{L_QL_i}}A_1\frac{L_{0}^{2}}{L_T}.
\end{align}
Consequently, the time-dependent interaction Hamiltonian is obtained.

By tuning the frequency of the ac part, $\Omega _{d}^{i}$, we can control the relative phase between the two coupling points. Specifically, when $\Omega _{d}^{i}=\Delta_i=\omega_e-\omega_i$, the modulation is resonant with the frequency detuning between the emitter and site $i$. For simplicity, we set $\omega_1=\omega_2$ and $g_1=g_2=g^{\prime}$. Under the condition $\Omega _{d}^{i}\gg g^{\prime}$, and in the rotating frame, the effective interaction Hamiltonian reduces to 
\begin{align} 
	 H_{\mathrm{int}} &=	\frac{g^{\prime}}{2}e^{-i\phi _1}\sigma _-a^{\dagger}+\frac{g^{\prime}}{2}e^{-i\phi _2}\sigma _-b^{\dagger}+\mathrm{H.c.}
	\\
&=g\sigma _-a^{\dagger}+ge^{i\Delta \phi}\sigma _-b^{\dagger}+\mathrm{H.c.} 
\end{align}
where $g=\frac{g^{\prime}}{2}$ denotes the equivalent coupling strength, while $\Delta \phi =\phi _1-\phi _2$ represents the relative phase between the two coupling sites.

%\bibliography{1D_ladder_reference}

\end{document}